\documentclass[aps,prl,reprint,superscriptaddress]{revtex4-2}
\usepackage{graphicx}
\usepackage{amsmath,amssymb,bm}
\usepackage{xcolor}
\newcommand{\EB}{\mathbf{E}\times\mathbf{B}}

\begin{document}

\title{Counter-rotating density and current structures in a partially magnetized
$\mathbf{E}\times\mathbf{B}$ plasma}

\author{Pengze Xiao}
\affiliation{School of Physics, Huazhong University of Science and Technology, Wuhan 430074, China}

\author{Ya Zhang}
\email{yazhang@whut.edu.cn}
\affiliation{Department of Physics, Wuhan University of Technology, Wuhan 430070, China}

\author{Hongyu Wang}
\affiliation{School of Physics, Anshan Normal University, Anshan 114007, Liaoning, China}

\author{Wei Jiang}
\email{weijiang@hust.edu.cn}
\affiliation{School of Physics, Huazhong University of Science and Technology, Wuhan 430074, China}

\author{Jean-Pierre Boeuf}
\email{jpb@laplace.univ-tlse.fr}
\affiliation{LAPLACE, Universit\'e de Toulouse, CNRS, INPT, UPS, 118 Route de Narbonne, 31062 Toulouse, France}

\begin{abstract}
The first 3D kinetic simulation of a magnetron discharge reproduces
the measured rotating spoke and reveals that density and current rotate in
opposite directions: the $m=1$ spoke turns in the $-\EB$ direction at
90~kHz while $m\simeq16$ electron-cyclotron-drift-type filaments turn at
1.0~MHz, with no net propagation along $\mathbf{B}$. Magnetic drifts
exchange far more energy than they deposit, yet their spoke-front heating
sustains the ionization. Following helical paths, the anode-directed
current is relayed by de-trapping turbulence at the sheath edge (50\%) and
by the spoke channel in the bulk (97\%).
\end{abstract}

\maketitle

Rotating ionization spokes are widely observed in partially magnetized $\EB$
plasmas, including magnetron discharges and Hall thrusters, where they
regulate ionization, power deposition, and anomalous cross-field electron
transport \cite{kaganovich2020,boeuf2023review,gudmundsson2022,roberts2024}. Coherent
Thomson scattering in a high-power pulsed planar magnetron has also
revealed millimetre-scale density fluctuations at MHz frequencies,
identified as the electron cyclotron drift instability (ECDI) and
amplitude-modulated at the kHz rate of the passing
spoke \cite{tsikata2015}. How these two very different scales combine to produce the anomalous
transport has remained an open question. Spoke-resolved probe and imaging measurements
have quantified the associated modulations of potential, electron energy,
density, and current \cite{held2020,held2022,maass2021,przybocki2024,panjan2024},
while kinetic and fluid studies connect the linear stage to gradient-drift or
Simon--Hoh-type instabilities and the nonlinear stage to ionization feedback
\cite{boeuf2013,boeuf2020,boeuf2023spoke,xu2021,xu2023,hara2022}. Most kinetic
descriptions, however, are two-dimensional: they suppress field-line curvature
and impose $k_\parallel=0$ by construction, leaving the roles of curvature and
parallel structure in electron heating and transport unresolved. A recent
$(z,\theta)$ kinetic study of a very similar configuration \cite{xu2026}
confirms the magnetic-gradient-drift heating at the spoke front.

In this Letter, we report a fully kinetic three-dimensional (3D) simulation
that reproduces the spoke-resolved measurements of Held
\emph{et al.}~\cite{held2022}, closes the discharge power balance, and
quantifies, channel by channel, the gross and net electron energization
by the magnetic-gradient and field-line-curvature drifts. It further
reveals that the density spoke and the fine-scale current pattern rotate
in opposite directions, and separates the cross-field flux, mode by mode,
into a coherent spoke-driven and a turbulent sheath-edge contribution.

A 3D electrostatic particle-in-cell/Monte Carlo collision (PIC/MCC) model with
implicit time integration is applied to the planar-magnetron geometry of Held
\emph{et al.}~\cite{held2022}. The domain is
$4\,\mathrm{cm}\times4\,\mathrm{cm}\times3\,\mathrm{cm}$ with a
$200\times200\times150$ uniform mesh ($\Delta x=0.2$~mm) and
$\Delta t=5\times10^{-11}$~s. Argon at 5~mTorr is sustained by a cathode
connected to a dc supply through a series resistor; the cathode voltage
stabilizes near $-260$~V. These conditions correspond to a low-current dcMS discharge, not to
HiPIMS, where the spoke rotation direction reverses at higher
current~\cite{yang2014,hecimovic2018}. The magnetostatic field of the experiment, electron-- and ion--neutral
collisions, and ion-induced secondary emission at the cathode are included;
the discharge is evolved for about 100~$\mu$s from a
$10^{12}\,\mathrm{m}^{-3}$ seed. Boundary conditions and sensitivity tests
are given in the Supplemental Material~\cite{suppl}.

\begin{figure}[!t]
\includegraphics[width=\columnwidth]{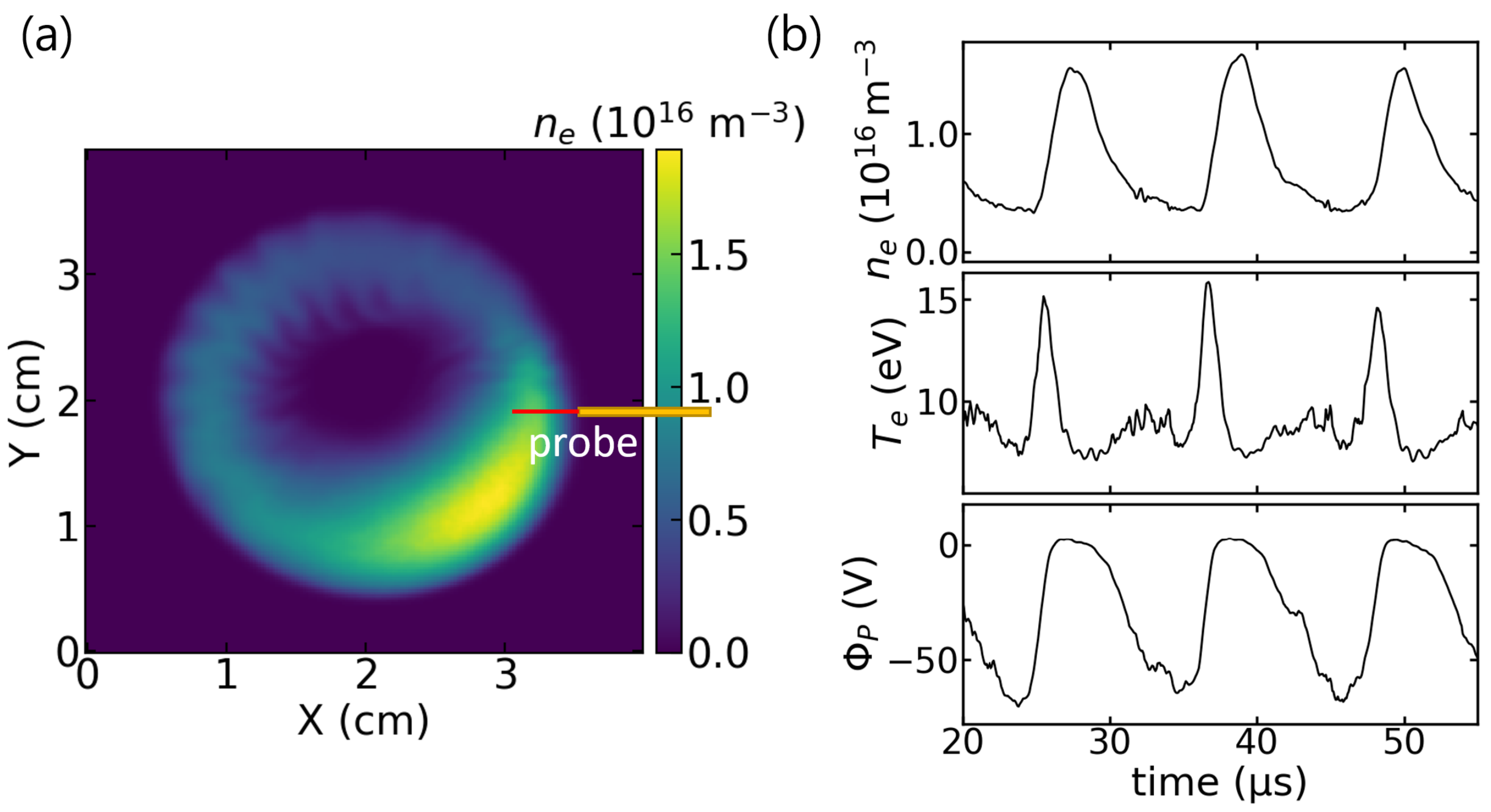}
\caption{Simulated spoke at the experimental probe position of Held
\emph{et al.}~\cite{held2022}. (a)~Electron-density snapshot in the racetrack
plane; the line indicates the numerical probe. (b)~Electron density $n_e$,
electron mean energy $T_e$, and plasma potential $\Phi_P$ at the probe. The
corresponding measured waveforms of Ref.~\cite{held2022} are reproduced in the
Supplemental Material~\cite{suppl}.}
\label{fig1}
\end{figure}

After the initial transient, an azimuthally localized potential structure
forms along the racetrack and develops into a saturated $m=1$ spoke.
Figure~\ref{fig1} shows its simulated passage at the experimental probe
position, for direct comparison with the spoke-resolved measurements of Held
\emph{et al.}~\cite{held2022,suppl}. In both the simulation and the
experiment, the spoke front is marked first by a rapid increase in plasma
potential and electron mean energy, followed by the maximum in electron
density, with comparable amplitudes; the sharper measured leading edge falls where
the probe-derived quantities are least reliable~\cite{held2022,suppl}. This ordering identifies the spoke front as a
localized electron-energization and ionization region --- the double
layer bounding the spoke in two-dimensional
models~\cite{boeuf2023spoke} --- rather than a passively
advected density perturbation. The 90~kHz rotation is three times the
$\simeq$30~kHz of Ref.~\cite{held2022}, at a simulated plasma density about
an order of magnitude below the experiment (the effective cathode
emission is poorly known, and the lower density also keeps the Debye
length and plasma period resolvable at the present mesh and time step);
the spoke velocity,
$\simeq$7~km\,s$^{-1}$, lies within the observed
1--10~km\,s$^{-1}$ range. The saturated density spoke propagates in the
$-\EB$ direction, consistent with the nonlinear ionization-wave regime found
in two-dimensional studies~\cite{boeuf2020,boeuf2023spoke} and with
the propagation direction observed experimentally in low-current dc
magnetrons~\cite{yang2014}. Fig.~\ref{fig1}(a) also reveals weak short-wavelength density corrugations
within the broad envelope; their connection to the high-$m$ current
filaments is examined below.

To identify the electron-energization mechanism, we decompose the work done on
a magnetized electron into classical collisional, magnetic-gradient, and
field-line-curvature contributions~\cite{boeuf2023spoke,boeuf2020}. With the
electron energies $\varepsilon_\perp$ and $\varepsilon_\parallel$ expressed in
eV, the heating rate per electron is $\Theta=-\mathbf{v}_d\cdot\mathbf{E}$ in
eV/s, and the corresponding power density is $p_e=en_e\Theta$. The three
contributions are
\begin{align}
\Theta_{\mathrm{coll}} &= \frac{\nu_m\Omega_{ce}}{\nu_m^2+\Omega_{ce}^2}\,
\frac{E_\perp^2}{B}\simeq\frac{\nu_m}{\Omega_{ce}}\frac{E_\perp^2}{B},
\label{eq:coll}\\
\Theta_{\nabla B} &= \frac{\varepsilon_\perp}{B}
\left(\mathbf{b}\times\frac{\nabla B}{B}\right)\cdot\mathbf{E},
\label{eq:gradB}\\
\Theta_{\mathrm{curv}} &= \frac{2\varepsilon_\parallel}{B}
\left[\mathbf{b}\times(\mathbf{b}\cdot\nabla)\mathbf{b}\right]\cdot\mathbf{E},
\label{eq:curv}
\end{align}
where $\mathbf{b}=\mathbf{B}/B$, $\nu_m$ is the electron--neutral
momentum-transfer frequency, and $\Omega_{ce}=eB/m_e$.
Equation~(\ref{eq:curv}) is intrinsically 3D: it vanishes from a planar model
that does not retain the curvature of the magnetic field lines.

\begin{figure}
\includegraphics[width=0.96\linewidth]{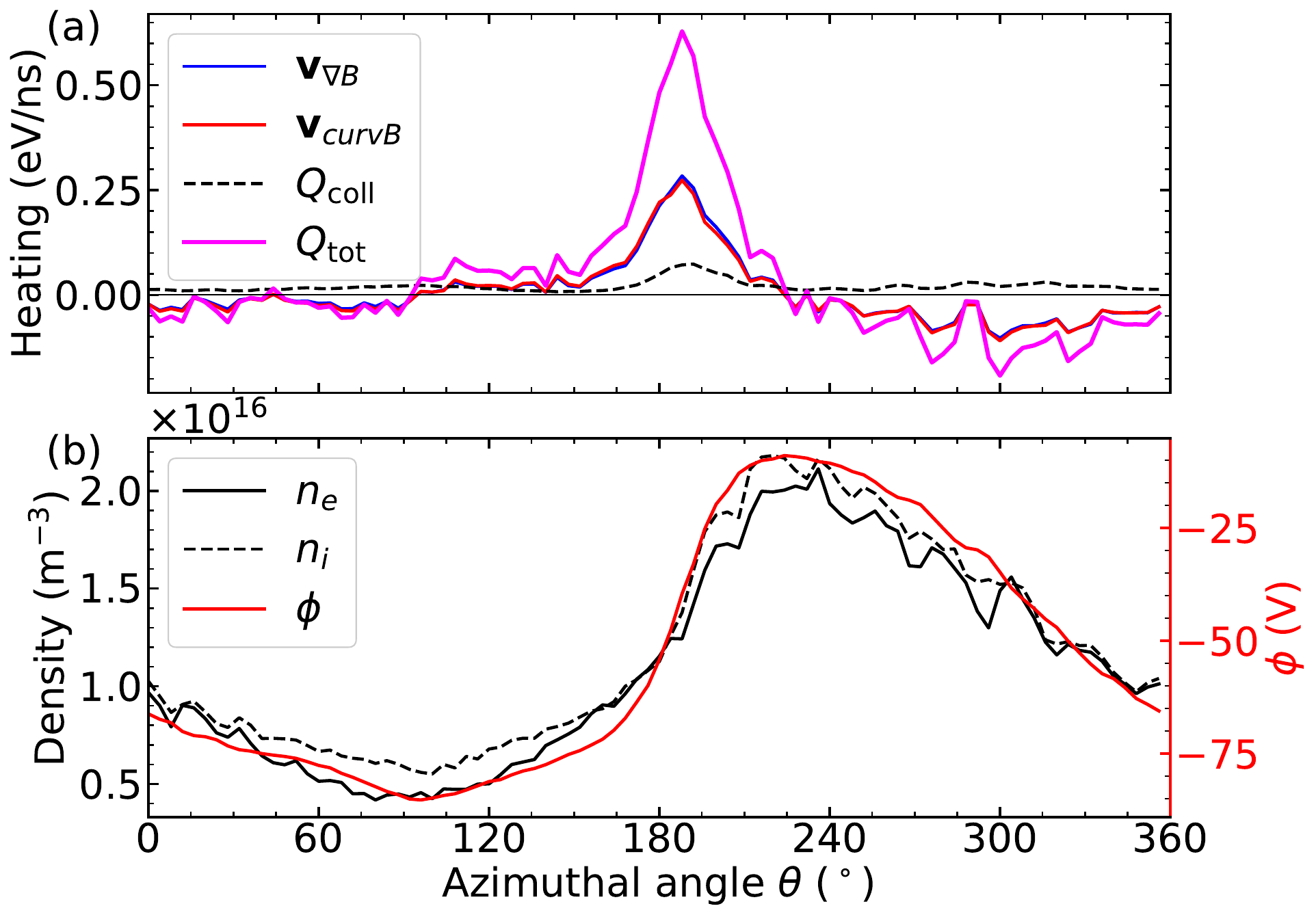}
\caption{Azimuthal profiles within a torus of major radius 1.2~cm and minor
radius 1~mm, 3~mm above the cathode. (a)~magnetic-gradient,
field-line-curvature, classical collisional, and total electron-heating rates
per electron. (b)~electron density, ion density, and plasma potential.}
\label{fig2}
\end{figure}

Figure~\ref{fig2}(a) shows that $\Theta_{\nabla B}$ and $\Theta_{\mathrm{curv}}$
possess nearly identical broad azimuthal envelopes: both become positive
across the sector where the plasma potential rises and the density then
increases, with comparable peak amplitudes, and both become negative
downstream. This is not a net energy loss: the same drifts that energize an
electron at the spoke front ($E_\theta>0$) take part of that energy back
behind it ($E_\theta<0$), and the cycle-integrated contribution remains
positive. Within the toroidal sampling volume of Fig.~\ref{fig2}(a), the
magnetic-gradient and curvature drifts account for 42.1\% and 42.3\% of
the drift-plus-collisional heating (collisions: 15.6\%, concentrated near
the sheath edge where $\Theta_{\mathrm{coll}}\propto E_\perp^{2}$ is
largest): locally the curvature term is of the same order as the
$\nabla B$ drift, not a perturbative correction to the two-dimensional
picture.
Integrated over the discharge volume and averaged over the
saturated phase, however, the positive and negative
drift contributions partially cancel: the $\nabla B$ channel exchanges
$+0.61$~W and $-0.45$~W for a net of $+0.16$~W (30.5\% of the net electron
heating), while the curvature channel exchanges $+0.46$~W and $-0.45$~W for a
net of only $+0.013$~W (2.6\%). Together the two drifts
account for 33.0\% of the total electron Joule heating
$\int\mathbf{J}_e\cdot\mathbf{E}\,dV$ computed from the simulation
fields, while classical collisions, whose heating rate
$\Theta_{E}\propto\nu E_\perp^{2}/B$ is always positive, contribute the
largest single net share (46\%). The remaining $\simeq$20\% lies
beyond these guiding-centre diagnostics. It points to the parallel
dynamics along the field lines and to the ballistic acceleration of
secondary electrons across the cathode sheath --- a direct
$\mathbf{J}_e\cdot\mathbf{E}$ channel that no drift describes --- rather
than to the fluctuations themselves, whose measured amplitude bounds
the power they can mediate to a small fraction of this residual (see
below). The curvature
drift is thus a nearly reversible energy-exchange
channel --- a gross-versus-net distinction that only a three-dimensional
model can make. Reversible does not mean inert: the electrons it
lifts above the ionization threshold at the spoke front ionize before
the cooling phase reclaims the energy. These guiding-centre channels are complementary diagnostics
rather than an exhaustive partition of the Joule heating; the global
closure is the agreement between the circuit power
$\langle UI\rangle_t$ and the total power given to the
particles~\cite{suppl}.

The net shares do not measure the functional role of each channel. The magnetic-drift heating is concentrated at the spoke front and increases with
the electron energy ($\Theta_{\nabla B}\propto\varepsilon_\perp$,
$\Theta_{\mathrm{curv}}\propto\varepsilon_\parallel$), so it preferentially
feeds the ionizing tail of the distribution --- 84\,\% of the local heating in the near-racetrack volume
[Fig.~\ref{fig2}(a)], consistent with the
temperature rise preceding the density maximum in Fig.~\ref{fig1}.
Collisional heating is azimuthally quasi-uniform and energy-independent: it
maintains the bulk, while the magnetic drifts sustain the ionization at the spoke
front. Because the filaments counter-rotate, their heating averages
out in the spoke frame within a microsecond; only the phase-locked
magnetic-drift heating provides the front--back asymmetry that organizes
the $m=1$ ionization wave.

The spatial structure of the energization is quantified by the
azimuthal spectrum of the electron-energy deposition $W_e$, whose $m$th
component measures the spatial modulation rather than a net heating
power:
\begin{equation}
W_{e,m}(r,z)=\frac{1}{2\pi}\int_{0}^{2\pi}
W_e(r,z,\theta)e^{-im\theta}\,d\theta .
\end{equation}
The resulting shares are $37.2\%$ ($m=0$), $35.3\%$
($m=1$--$3$), $5.8\%$ ($m=4$--$11$), $18.0\%$
($m=12$--$22$), and $3.8\%$ ($m>22$)~\cite{suppl}.


\begin{figure}
\centering
\includegraphics[width=1.0\linewidth]{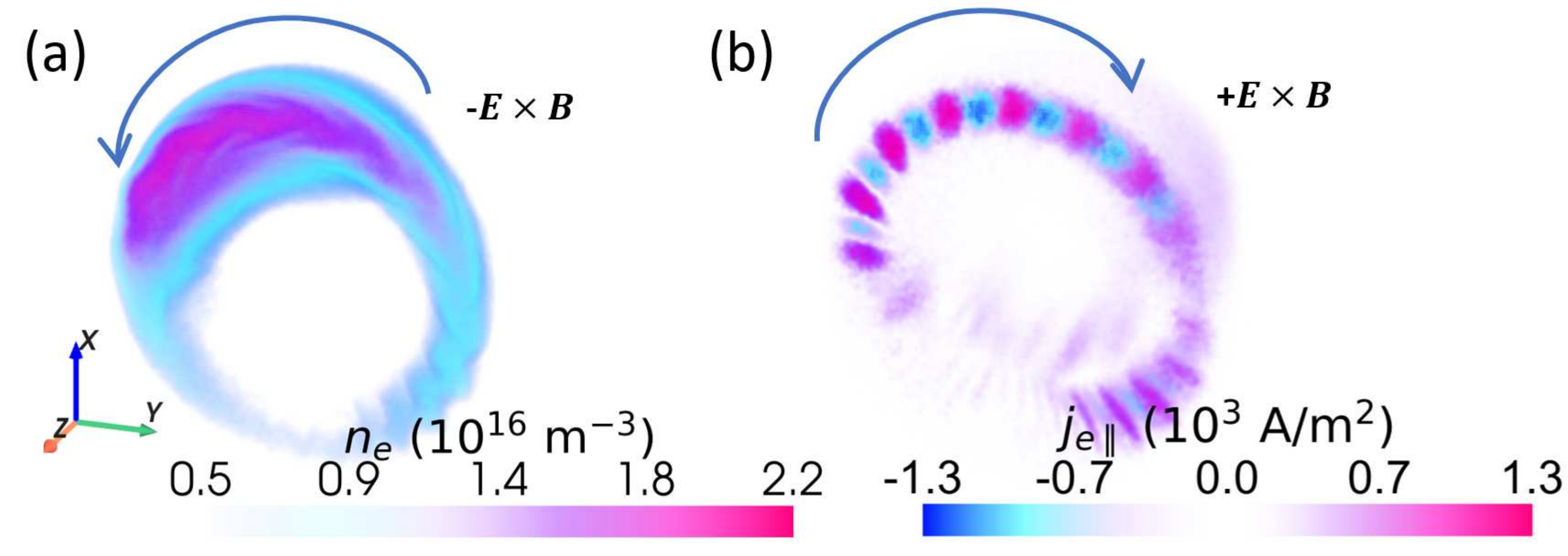}
\\[1.0ex]
\includegraphics[width=1.0\linewidth]{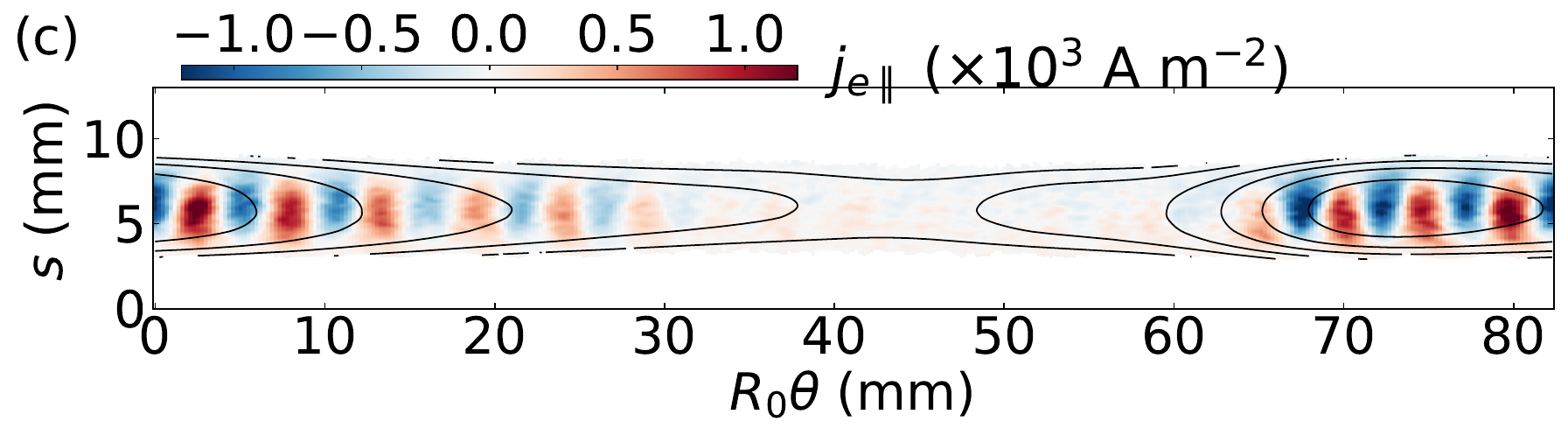}
\\[0.5ex]
\includegraphics[width=1.0\linewidth]{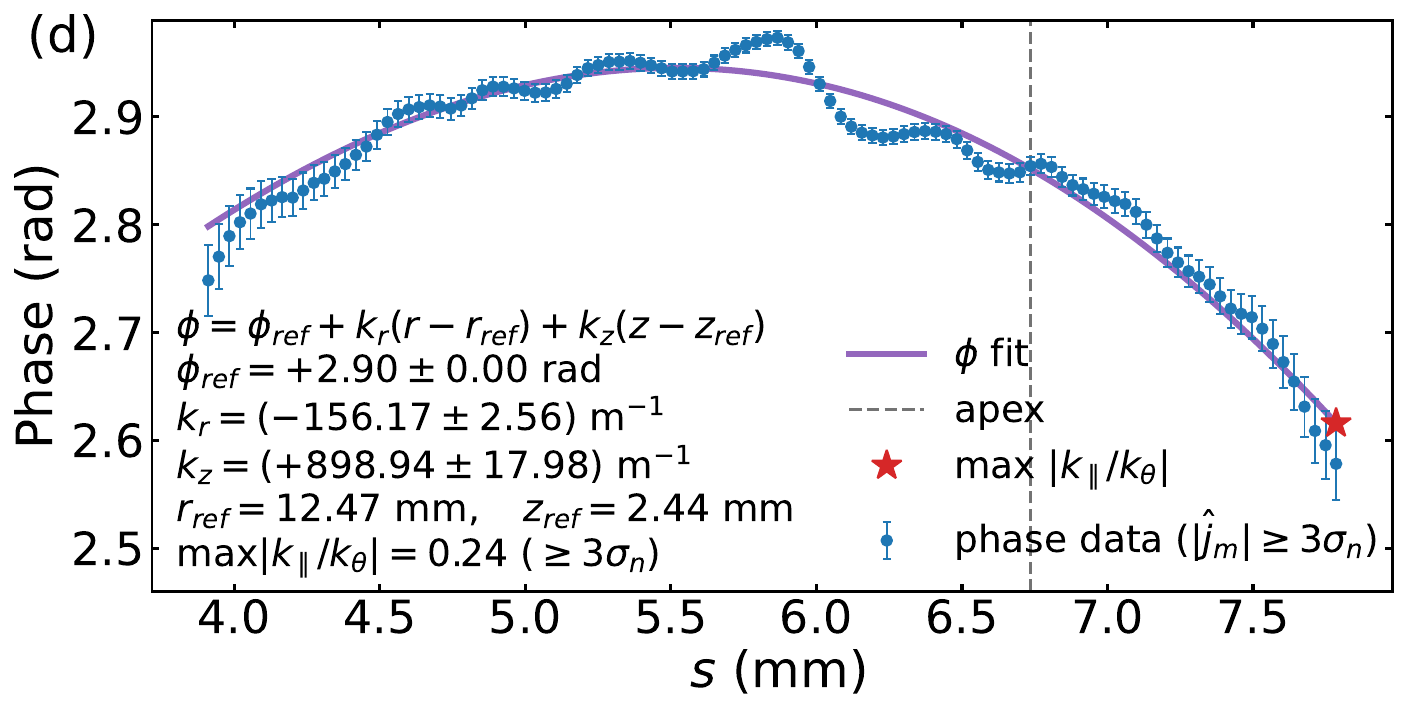}
\caption{(a)~Three-dimensional electron-density envelope: the broad $m=1$
spoke, propagating in the $-\EB$ direction. (b)~Field-aligned current
structure: the $m\simeq16$ filament pattern, propagating in the $+\EB$
direction; see also the video in the Supplemental
Material~\cite{suppl}. (c)~Field-aligned electron current
$j_{e\parallel}(s,R_0\theta)$ on the racetrack magnetic flux surface, in
magnetic coordinates with equal aspect ratio, with several equipotential contours of the low-$m$ potential overlaid.
(d)~Phase of the dominant azimuthal component of $j_{e\parallel}$ along
the field line; the curve is the two-parameter fit
$\varphi=\varphi_{\mathrm{ref}}+k_r\,(r-r_{\mathrm{ref}})+k_z\,(z-z_{\mathrm{ref}})$.
The potential and the electron temperature along the field line are
shown in the Supplemental Material~\cite{suppl}.}
\label{fig3}
\end{figure}

Figure~\ref{fig2}(b) reveals a feature hidden by the broad spoke-scale
profiles. Although the electron and ion densities share the same $m=1$
envelope, the electron density carries pronounced short-wavelength azimuthal
corrugations that the ion density does not follow. The resulting local
mismatch between $n_e$ and $n_i$ indicates finite charge separation at the
sheath edge, and therefore small-scale electrostatic fields superimposed on
the spoke potential, of amplitude $\delta E_\theta\simeq0.67$~kV/m
and relative density modulation $\delta n_e/n_e\simeq12.7\%$ in the
$12\leq m\leq22$ band at the sheath edge --- the signature of a
charge-separating high-$m$
instability embedded within the saturated spoke. Its mode number and parallel
structure are examined below using the field-aligned current. The electron potential energy
forms a $\simeq200$~eV deep well with its minimum near the apex
(Supplemental Material~\cite{suppl}), so
electrons shuttle back and forth along the arch, and the field-aligned
current arises from this electrostatically restored motion.
The parallel channel is itself a net sink: $J_\parallel E_\parallel$
exchanges $+2.02$~W and $-2.32$~W for a net of $-0.31$~W --- electrons
climbing out of the well return energy to the field. The field line is thus neither
isopotential nor isothermal --- $T_e$ varies strongly along the arch, with
a finite $T_\parallel$--$T_\perp$ anisotropy~\cite{suppl} --- in line
with recent Thomson-scattering evidence in a Hall
thruster~\cite{suazo2024}.


The two structures are shown side by side in Fig.~\ref{fig3}(a,b): the
field-aligned electron current on the racetrack flux surface
[Fig.~\ref{fig3}(c)] is dominated by a sequence of narrow filaments with
$m\simeq16$, whereas the density is dominated by the broad $m=1$ spoke.  Phase tracking
(video in the Supplemental Material~\cite{suppl}) shows that the
density spoke propagates in the $-\EB$ direction at $90.0$~kHz, whereas
the high-$m$ current pattern propagates in the $+\EB$ direction at
$1.0$~MHz. Density and current thus remain coupled through the same spoke-scale
potential envelope but are not two diagnostics of one rigidly rotating
mode. The same distinction separates density-resolved spoke
measurements~\cite{held2022} from the current oscillations observed in a
much smaller magnetron~\cite{przybocki2024}: the common point is the
structure, not the frequencies, which scale with each device. With the
racetrack radius $R_0=1.31$~cm, $m\simeq16$ corresponds to
$\lambda_\theta\simeq5.1$~mm
($k_\theta\simeq1.2\times10^{3}$~rad\,m$^{-1}$). The filaments extend over the whole azimuth, with $j_\parallel$
more pronounced in the high-density region --- the combined effect of the
larger electron population and the sheath field.

These properties identify the filaments as the three-dimensional
counterpart of the ECDI detected by coherent Thomson scattering in a
planar magnetron~\cite{tsikata2015}. The experiment operates in the far
denser HiPIMS regime, so the comparison is made in dimensionless terms
($k\lambda_{De}$, $v_{\mathrm{ph}}/c_s$), not in absolute scales. The
filament frequency, $1.0$~MHz, is measured directly from the temporal
phase of the $m=16$ component. The phase velocity,
$2\pi f/k_\theta\simeq5.2$~km\,s$^{-1}$, is within 10\% of the ion
sound speed at the $\simeq10$~eV electron temperature, with
$k\lambda_{De}\simeq0.3$: this is the long-wavelength, ion-acoustic-like
regime into which the saturated ECDI is known to
evolve~\cite{lafleur2016}, and which the
Thomson spectrum resolves as a continuous dispersion. The discrete ECDI
resonances themselves, $\lambda\lesssim2\pi v_E/\Omega_{ce}\sim1$~mm,
would span only a few cells of the 0.2~mm mesh and cannot be resolved
here. The filaments propagate in the $+\EB$ direction, carry the charge
separation of Fig.~\ref{fig2}(b), and have an amplitude modulated at
the $90$~kHz rate of the passing spoke. The measured amplitudes bound the power these fluctuations
can mediate: the free energy stored in the density corrugation,
$\sim n_eT_e(\delta n_e/n_e)^2$, exchanged even at the wave frequency,
yields at most a few mW --- the fluctuations relay the current, not the
power. A quantitative discrimination between the ECDI and
the oblique modified two-stream instability
(MTSI)~\cite{petronio2021,janhunen2018} requires comparing the measured
$(\omega,k_\theta,k_\parallel)$ with the two linear dispersion
relations; the field-aligned structure needed for that comparison is
quantified below. The quantitative results of this Letter --- the
band-resolved transport, the heating channels, and the $k_\parallel$
bound --- involve only scales $m\leq22$
($\lambda_\theta\geq3.7$~mm, at least 18 cells) and do not depend on
this identification.

The two structures are therefore dynamically coupled --- the
three-dimensional counterpart of the kHz modulation of the MHz-scale
scattered signal of Ref.~\cite{tsikata2015}.

The magnetic-coordinate map of Fig.~\ref{fig3}(c) separates the parallel
structure of the filaments from a geometrical projection: on an
equal-aspect $(s,R_0\theta)$ map, the stripe tilt $\alpha$ from the
field-line direction measures $k_\parallel/k_\theta=\tan\alpha$
($k_\theta=m/R_0$), but the angle is too small to be read reliably. We
extract $k_\parallel$ instead from the phase of the dominant azimuthal
component of the field-aligned current along the field line,
$\hat{j}_{\parallel,m}(s)$ [Fig.~\ref{fig3}(d)]. The phase increases from the cathode footpoint toward a maximum located
inside the arch and decreases beyond it, so its slope changes sign: there is no
net propagation along $\mathbf{B}$. What the profile measures is the local
field-aligned projection $k_\parallel(s)=\mathbf{k}\cdot\hat{b}(s)$ of a
wave vector of fixed orientation; the projection vanishes where
$\mathbf{k}\perp\mathbf{B}$ and is bounded by
$|k_\parallel|/k_\theta\leq0.24$ over the measured band
($|\hat{j}_{\parallel,m}|\geq3\sigma$). The two-parameter fit of
the phase, $\varphi(s)=\varphi_{\mathrm{ref}}+k_r\,[r(s)-r_{\mathrm{ref}}]
+k_z\,[z(s)-z_{\mathrm{ref}}]$, gives
$k_r=-156\pm3~\mathrm{m}^{-1}$ and $k_z=+899\pm18~\mathrm{m}^{-1}$: the
filaments carry fixed radial and vertical wave-number components set by the
discharge geometry, i.e.\ wavefronts tilted in the $(\theta,z)$ plane
($k_z/k_\theta\simeq0.74$, vertical wavelength $\simeq7$~mm); the
corresponding stripe tilt stays below $\sim13^{\circ}$, too small to read
off the map. The
high-$m$ component is therefore quasi-perpendicular, with a small oblique
admixture that reverses along the magnetic arch. This is consistent with the electrostatic shuttling above: the
current oscillates without net parallel propagation, with a spatial phase
fixed by $(k_r,k_z)$.
This measurement has no two-dimensional counterpart --- a $(z,\theta)$ or
$(r,\theta)$ model imposes $k_\parallel=0$ by construction~\cite{xu2026} ---
and it is what linear theory needs but experiment cannot supply: a
field-aligned projection that varies along the arch smooths the discrete
cyclotron resonances into the continuous dispersion relation measured in the
magnetron~\cite{tsikata2015}.
\begin{figure}
\centering
\includegraphics[width=0.9\linewidth]{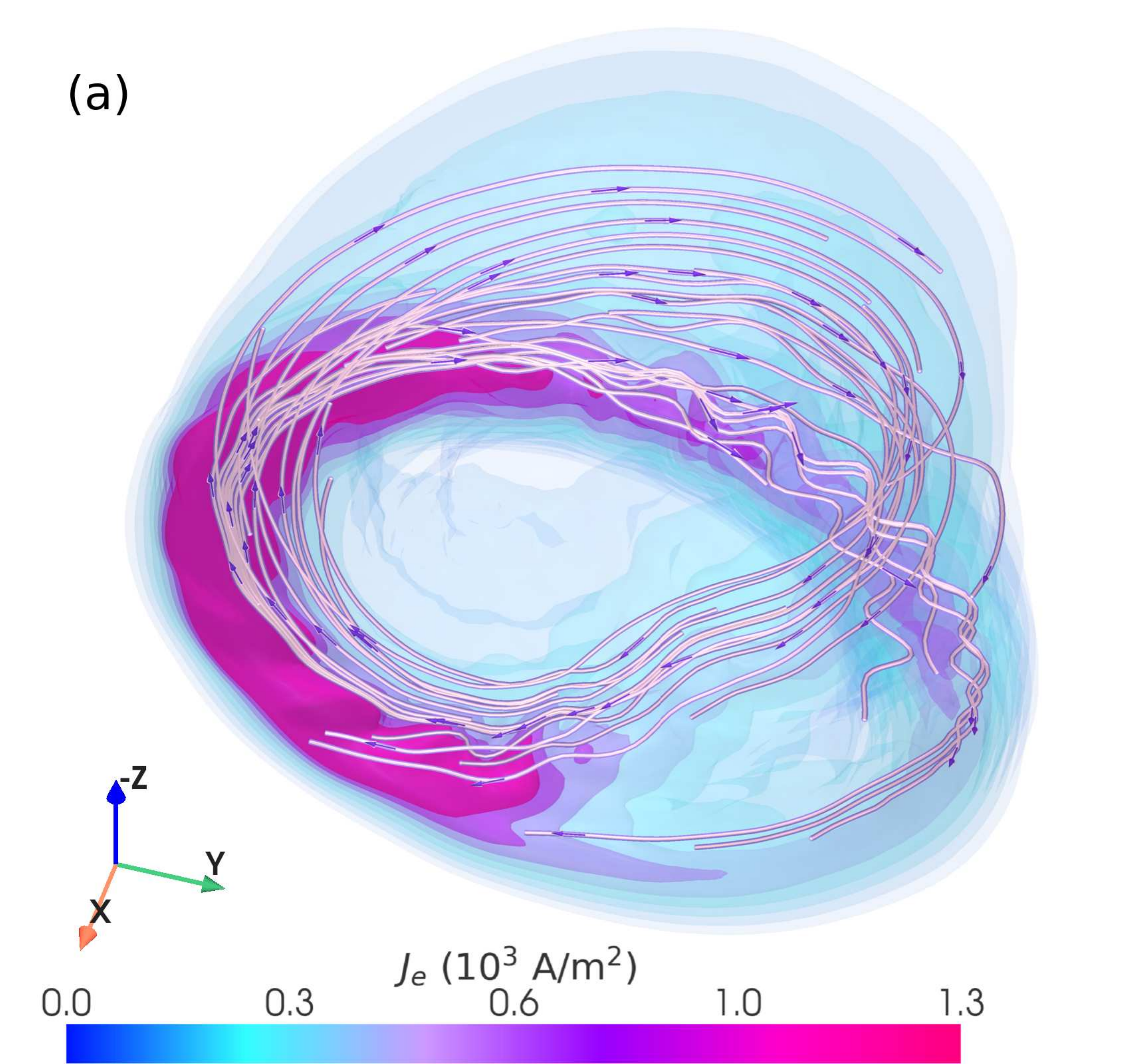}
\\[2.5ex]
\includegraphics[width=0.9\linewidth]{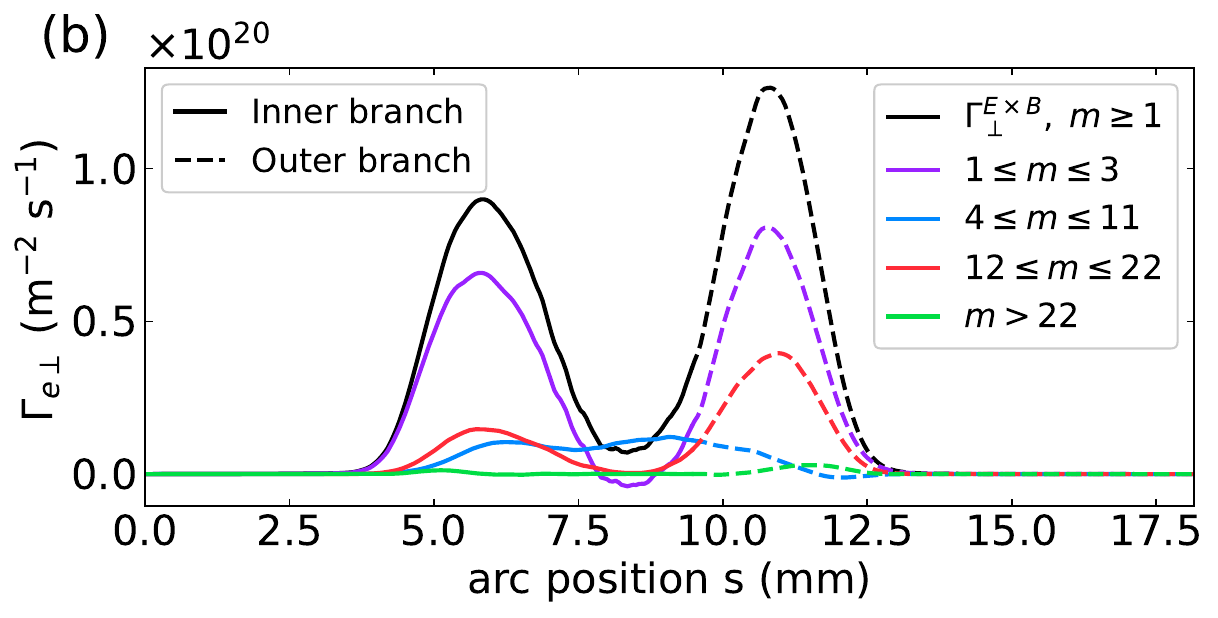}
\caption{(a)~Three-dimensional electron-current structure: current-density
magnitude (colormap, truncated at the 99.6th percentile of all grid-cell
values) and representative helical current streamlines, with arrows
indicating the electron motion. (b)~Band-integrated mode-resolved
electron cross-field flux $\Gamma_\perp$ [Eq.~(\ref{eq:gamma})], positive
toward the anode, against arc position along the flux surface, from the
cathode footpoint to the apex, for the inner (solid)
and outer (dashed) branches of the magnetic flux surface: azimuthal-mode
bands and their sum over all $m\geq1$. The profile shows
\emph{where} the surface is crossed; the net anode-directed transport is
the surface-integrated flux.}
\label{fig4}
\end{figure}

Figure~\ref{fig4} connects the counter-rotating patterns to the cross-field
transport. The electron-current streamlines form helical paths through the curved
magnetic geometry and concentrate where the spoke deforms the plasma
potential, whose equipotential surfaces extend from the cathode side toward
the anode --- the three-dimensional counterpart of the spoke-induced
equipotential ``short circuit'' identified in two
dimensions~\cite{boeuf2023spoke}.

Decomposing the electron density and the azimuthal field into
azimuthal Fourier components, $n_m(z,t)$ and $E_{\theta,m}(z,t)$, on the
racetrack flux surface --- flux line 2 defined below,
$(r,z)=(12~\mathrm{mm},\,5~\mathrm{mm})$ --- the contribution of mode
$m$ to the electron flux
across $\mathbf{B}$ is
\begin{equation}
\Gamma_\perp(m,z)=\frac{2}{B(z)}\,
\Big\langle\,\mathrm{Re}\big[\,n_m(z,t)\,E^{*}_{\theta,m}(z,t)\,\big]\Big\rangle_t,
\label{eq:gamma}
\end{equation}
where $\langle\cdot\rangle_t$ is a time average over the saturated phase.
Summed over all $m\geq1$, Eq.~(\ref{eq:gamma}) reproduces the cross-field
flux obtained independently by counting macroparticles crossing the same
surface to within 18.1\,\%, confirming that the anomalous
cross-field transport is overwhelmingly of $\EB$ nature. The residual
is numerical: the counting surface, built from grid cells, follows the
field line only approximately, and wherever it deviates part of the
fast parallel shuttle motion is counted as cross-field flux --- most
strongly near the apex, where the shuttle is fastest and the two
branches join. The truly
collisional contribution is bounded by $\nu_m/\Omega_{ce}\ll 1$.


Three flux lines at increasing height above the cathode are compared:
$(r,z)=(12,\,2.6)$~mm (line 1, near-sheath --- also the line of the
field-aligned analysis of Fig.~\ref{fig3}), $(12,\,5)$~mm (line 2, the
reference case of Fig.~\ref{fig4}(b)), and $(12,\,9)$~mm (line 3, bulk);
their layout and the full mode table are given in the Supplemental
Material~\cite{suppl}.

Integrated along flux line 2, the coherent spoke band $1\leq m\leq 3$
carries 63.7\,\% of the net cross-field flux and the filamentary
turbulent band $12\leq m\leq 22$ contributes 23.2\,\%. The intermediate
band $4\leq m\leq 11$ (12.0\,\%) is the azimuthal tail of the spoke:
under the $-\EB$ rotation, electrons are deflected away from the anode at
the leading edge and toward it at the trailing edge, a net positive
transport that places this band in the coherent spoke channel
(75.7\,\% in total for $m\leq11$); modes with $m>22$ contribute
1.1\,\%.

Moving from the sheath into the bulk, the dominant band crosses
over: on the near-sheath line 1 the turbulent band $12\leq m\leq22$
carries 50.0\,\% of the flux, while on the bulk line 3 the spoke band
$1\leq m\leq3$ carries 96.7\,\%. The filaments transport little by
themselves --- their alternating sign cancels in a full azimuthal
integral --- but at the sheath edge they act as the de-trapping mechanism,
releasing electrons from the magnetic trap where the spoke channel then
carries them to the anode.



In summary, the 3D kinetic simulation reproduces the measured rotating
spoke and closes its energy and transport budgets. The magnetic-gradient and
curvature drifts exchange large powers that nearly cancel (net
33.0\,\%); collisions contribute the largest single net share, but the
spoke-front magnetic-drift heating sustains the ionization. The same discharge sustains a broad $m=1$ density spoke and a
counter-rotating $m\simeq16$ current pattern of
electron-cyclotron-drift type,
quasi-perpendicular and with no net propagation along $\mathbf{B}$. A mode-resolved decomposition shows that the spoke and its azimuthal
tail carry 75.7\,\% of the anode-directed flux on the reference surface,
while the share of the sheath-edge turbulence rises from 3.7\,\% in the
bulk to 50\,\% at the sheath edge: the two contributions exchange their
dominant role at the sheath--bulk interface.

Density propagation, current-pattern propagation, and net cross-field
transport are therefore three distinct but coupled elements of the nonlinear
spoke, which sustains the small-scale turbulence and is in turn fed by
it. This explains the coexistence of
kHz and MHz scales observed experimentally in magnetron
discharges~\cite{tsikata2015} and shows that magnetic-field curvature and
parallel structure are essential ingredients of the energy exchange and
anomalous transport in partially magnetized plasmas.

\begin{acknowledgments}
This work was supported by the National Natural Science Foundation of China under Grants No.~12275095, No.~12675325, and No.~W2621001.
\end{acknowledgments}

\end{document}


\title{\texorpdfstring{Supplemental Material for:\\
Counter-rotating density and current structures in a partially magnetized
$E\times B$ plasma}{Supplemental Material for: Counter-rotating density and current structures in a partially magnetized E x B plasma}}

\author{Pengze Xiao}
\affiliation{School of Physics, Huazhong University of Science and Technology, Wuhan 430074, China}

\author{Ya Zhang}
\email{yazhang@whut.edu.cn}
\affiliation{Department of Physics, Wuhan University of Technology, Wuhan 430070, China}
\author{Hongyu Wang}
\affiliation{School of Physics, Anshan Normal University, Anshan 114007, Liaoning, China}

\author{Wei Jiang}
\email{weijiang@hust.edu.cn}
\affiliation{School of Physics, Huazhong University of Science and Technology, Wuhan 430074, China}

\author{Jean-Pierre Boeuf}
\email{jpb@laplace.univ-tlse.fr}
\affiliation{LAPLACE, Universit\'e de Toulouse, CNRS, INPT, UPS, 118 Route de Narbonne, 31062 Toulouse, France}

\maketitle

\section{Comparison with spoke-resolved measurements}

\begin{figure}[!htb]
  \centering
  \includegraphics[width=0.98\textwidth]{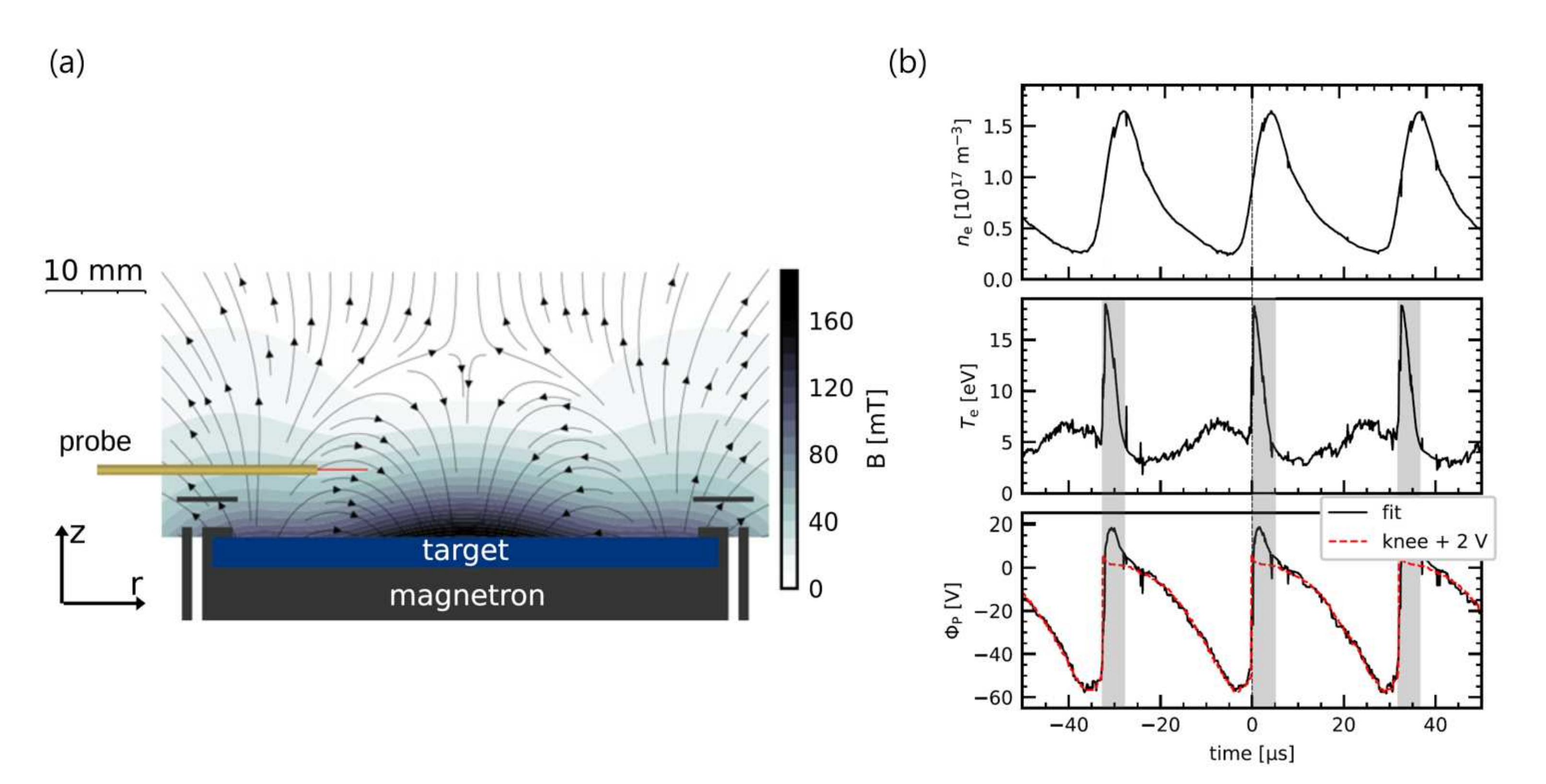}
  \caption{\label{fig:probe-comparison}
  Comparison with the spoke-resolved probe measurement.
  (a) Cross section of the magnetron geometry showing magnetic-field lines and
  the probe position. The figure is true to scale. Adapted from Held et al 2020 Plasma Sources Sci. Technol. 29 125003 \cite{Held2020}. Copyright 2020 Authors, licensed under a Creative
Commons Attribution (CC BY) license. 
  (b) Corresponding experimental waveforms from Ref.~\cite{Held2022}.  The
  shaded bands mark intervals in which the probe-derived quantities are least
  reliable; the red dashed curve denotes the fitted potential knee shifted by
  $2~\mathrm{V}$. Adapted from Held et al 2022 Plasma Sources Sci. Technol. 31, 085013 \cite{Held2022}. Copyright 2022 Authors, licensed under a Creative
Commons Attribution (CC BY) license. 
  }
\end{figure}

The numerical probe is positioned at the identical radial and axial coordinates as those of the experimental probe in Held \textit{et al.}~\cite{Held2022}. Specifically, it is located at a radial distance of approximately 20 mm from the target center and at an axial distance of 6 mm from the cathode surface; the probe itself has a length of 5 mm. The geometry and the probe are shown in
Fig.~\ref{fig:probe-comparison}(a). 

The measured waveforms shown in
Fig.~\ref{fig:probe-comparison}(b) are compared with the simulated waveforms presented in Fig.~1(b) of the Letter.  Both exhibit the same
ordering at the spoke front: the plasma potential and electron mean
energy rise first, followed by the maximum in electron density.  The
comparison is therefore based on the relative phase and modulation of
the three quantities, rather than on an arbitrary alignment of their
time origins.  The sharpest part of the measured leading edge lies
within the shaded intervals, where the probe-derived quantities are
least reliable.

\section{Numerical model and convergence}

To investigate the formation and dynamics of rotating spokes, we developed a fully three-dimensional implicit particle-in-cell/Monte Carlo collision (PIC/MCC) model accelerated on graphics processing units (GPUs). In contrast to many previous three-dimensional kinetic simulations\cite{taccogna2018,villafana2023}, the present model retains the experimental system size and the relevant physical temporal and spatial scales without applying artificial geometric, mass-ratio, or time-scale rescaling. This approach avoids scaling-induced numerical artifacts and preserves the physical hierarchy of the characteristic plasma time and length scales.

Compared with conventional explicit PIC schemes, implicit PIC methods provide substantially greater flexibility in the choice of numerical time step and spatial resolution\cite{lapenta2012}. In explicit PIC simulations, numerical stability generally requires the electron plasma oscillation and other fast electron time scales to be sufficiently resolved, while the grid spacing is typically constrained by the Debye length to avoid numerical instabilities and excessive numerical heating. These requirements become particularly restrictive in large-scale, multidimensional simulations, where the characteristic electron scales can be orders of magnitude smaller than the macroscopic spatial and temporal scales of interest. In an implicit formulation, the particle motion and field evolution are coupled self-consistently over each time step, which relaxes the stringent stability constraints imposed on explicit schemes and permits the use of larger time steps and grid sizes when the corresponding high-frequency and small-scale dynamics are not of primary interest. Consequently, implicit PIC enables efficient simulations over experimentally relevant system sizes and sufficiently long physical times while retaining the kinetic description of the plasma, making it particularly suitable for investigating the formation and long-time evolution of rotating spoke structures.

\begin{figure}[H]
  \centering
  \includegraphics[width=0.98\textwidth]{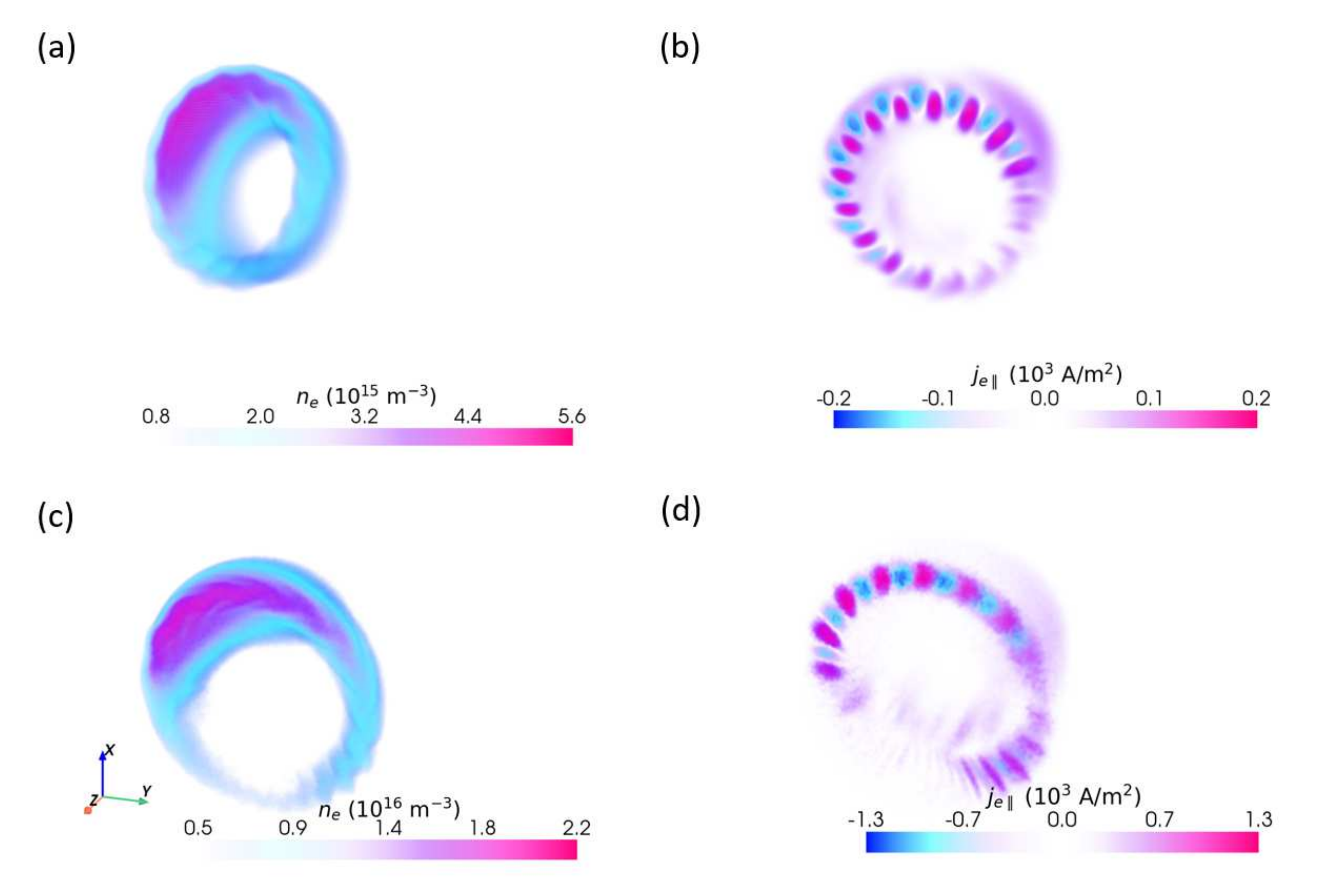}
  \caption{\label{fig:numerical-model}
  Convergence test: coarse-grid (a,b) vs. fine-grid (c,d) results for electron density (a,c) and field-aligned current (b,d). The fine grid is used for the main simulations.}
\end{figure}

The PIC module advances the trajectories of electrons and $\mathrm{Ar}^{+}$ ions on a three-dimensional computational grid. At each time step, the particles are first advanced to intermediate
positions, from which the predicted charge density
$\tilde{\rho}^{\,n+1}$ and the plasma susceptibility tensor
$\boldsymbol{\chi}$ are deposited onto the grid. The electrostatic
potential is then obtained self-consistently from the generalized
Poisson equation of the direct implicit PIC scheme,
\begin{equation}
\nabla\cdot
\left[
\varepsilon_{0}
\left(\mathbf{I}+\boldsymbol{\chi}\right)
\cdot\nabla\Phi^{n+1}
\right]
=
-\tilde{\rho}^{\,n+1},
\end{equation}
where $\varepsilon_{0}$ is the vacuum permittivity and
$\boldsymbol{\chi}$ accounts for the implicit plasma response.
Equivalently, the effective dielectric tensor is
$\boldsymbol{\varepsilon}
=\varepsilon_{0}(\mathbf{I}+\boldsymbol{\chi})$.
The electric field is subsequently evaluated as
$\mathbf{E}=-\nabla\Phi$, and the particle velocities and positions
are updated according to the Lorentz force. The use of an implicit time-integration scheme improves numerical stability while retaining the kinetic response required to describe the spoke and the associated small-scale current structures\cite{Wang2010}.


Collisions between charged particles and the neutral argon background are treated using the MCC method. The included electron--neutral processes comprise elastic scattering, excitation, and ionization, while the relevant ion--neutral collision processes are also taken into account. Collision events are selected probabilistically from their energy-dependent cross sections\cite{LXCat2016,Pitchford2017}, allowing the model to describe collisional momentum transfer, energy relaxation, and plasma production self-consistently.

Plasma--surface interactions are included because they play an essential role in discharge ignition and sustainment. Electrons incident on material boundaries are treated through three distinct channels: absorption, elastic reflection, and true-secondary-electron emission \cite{stoltz2003,Phelps1999}. Ion-induced secondary-electron emission is also included using an emission coefficient that depends on the local plasma density: it is initially set to 1.0 to promote rapid ionization; as the density rises, the coefficient decreases, and by the onset of spoke formation it returns to a typical value\cite{Safi2000} of 0.1. This treatment provides a self-consistent source of energetic secondary electrons and is more physically realistic than imposing a prescribed, time-independent electron beam.

To simulate the realistic electrical circuit, the anode is grounded while the cathode is placed in series with a $22\,\mathrm{k}\Omega$ resistor and a $-500\,$V power supply, resulting in a cathode voltage around $-260\,$V, in agreement with the experimental value.

Two sets of 3D electrostatic particle‑in‑cell/Monte‑Carlo‑collision simulations are performed for reference.
The coarse‑grid calculation adopts a $6~\mathrm{cm}\times6~\mathrm{cm}\times3~\mathrm{cm}$ domain discretized on a uniform $200\times200\times100$ mesh, yielding a cell size of $\Delta x=0.3~\mathrm{mm}$ and a time step $\Delta t=1\times10^{-10}~\mathrm{s}$.
The fine‑grid reference calculation is run on a $4~\mathrm{cm}\times4~\mathrm{cm}\times3~\mathrm{cm}$ domain with a uniform $200\times200\times150$ mesh, corresponding to $\Delta x=0.2~\mathrm{mm}$ and time step $\Delta t=5\times10^{-11}~\mathrm{s}$. The magnetic field, neutral pressure,
external circuit, collision processes, and secondary-electron-emission model
are identical to those used for the results in the Letter.

As shown in Fig.~\ref{fig:numerical-model}, despite the different domain extents (and hence boundary positions), the dominant azimuthal mode numbers of both the spoke and the small-scale current structures stay the same for the two combinations of grid resolution and time step. This insensitivity indicates that numerical discretization errors do not affect our main physical conclusions. We therefore adopt the finer grid and smaller time step for the results presented in the main text, allowing a more quantitative comparison with the experiment. The good agreement between the simulated and measured trends further supports the reliability of the three-dimensional model.

\section{Validation of the electron-heating diagnostic}
\begin{figure}[H]
  \centering
  \includegraphics[width=0.92\textwidth]
  {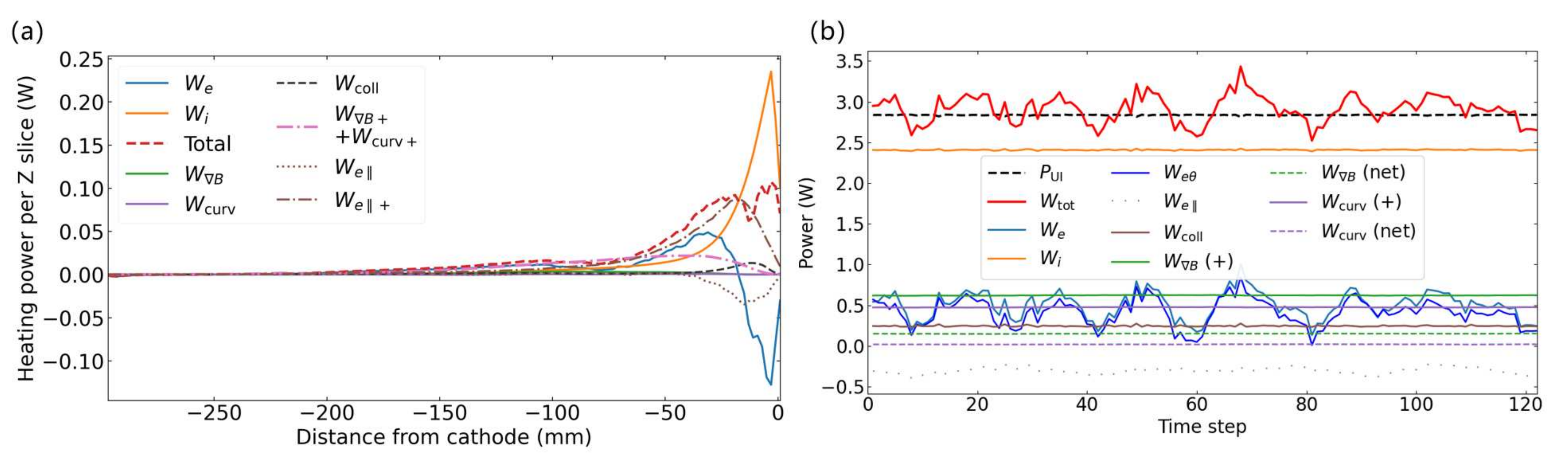}
  \caption{\label{fig:power-balance}
  Spatially resolved field--particle energy conversion and global discharge power balance.
(a) Axial distributions of the electron power 
$W_e(z)=\int_{\Delta V_z}\mathbf{J}_e\cdot\mathbf{E}\,{\rm d}V$, 
the ion power 
$W_i(z)=\int_{\Delta V_z}\mathbf{J}_i\cdot\mathbf{E}\,{\rm d}V$, 
their sum, the signed grad-$B$- and curvature-drift powers, 
the parallel electron power $W_{e\parallel}$ together with 
its positive component $W_{e\parallel,+}$, the collisional 
heating, and the sum of the positive drift contributions 
$W_{\nabla B}^{(+)}+W_{\mathrm{curv}}^{(+)}$.
  The difference between the positive and signed drift powers
  demonstrates the cancellation between local electron heating and
  cooling. The negative electron power immediately in front of the
  cathode results from cathode-directed electrons returning more
  energy to the electric field than is gained locally by emitted
  secondary electrons.
  (b) Electrical input power $P_{UI}=UI$, total particle power
  $W_{\mathrm{tot}}=W_e+W_i$, electron and ion powers, azimuthal and
  field-aligned electron work, collisional heating, and the positive
  and net grad-$B$- and curvature-drift contributions. Relative to the
  time-averaged electron power, the net grad-$B$ drift, net curvature
  drift, and collisional contributions are $30.4\%$, $2.6\%$, and
  $46.4\%$, respectively. The large difference between the positive
  and net curvature-drift powers indicates nearly complete
  cancellation between heating and cooling. The ion power varies only
  weakly, whereas the small-scale instability produces strong
  fluctuations of the net electron power. The agreement between the
  time-averaged $P_{UI}$ and $W_{\mathrm{tot}}$ demonstrates global
  power closure.
  }
\end{figure}

Figure~\ref{fig:power-balance} provides a spatially resolved view of
field--particle energy conversion together with a global validation of
the discharge power balance. The instantaneous electrical power
transferred from the field to species $s$ is defined as
\begin{equation}
W_s(t)
=
\int_V
\mathbf{J}_s(\mathbf{x},t)\cdot
\mathbf{E}(\mathbf{x},t)\,
{\rm d}V,
\qquad s=e,i ,
\end{equation}
where $W_s>0$ denotes energy transfer from the electric field to the
particles. The corresponding axial profiles in
Fig.~\ref{fig:power-balance}(a) are obtained by applying the same
integral to each $z$ slice,
\begin{equation}
W_s(z,t)
=
\int_{\Delta V_z}
\mathbf{J}_s\cdot\mathbf{E}\,
{\rm d}V .
\end{equation}
Most of the ion power is deposited within the cathode sheath, where the
ions are accelerated toward the cathode. The electron power has a more
complicated spatial structure and becomes negative immediately in
front of the cathode. Although secondary electrons emitted from the
cathode gain energy from the sheath electric field, this positive
contribution is locally exceeded by the energy returned to the field
by cathode-directed electrons.

The strongly nonuniform magnetic field controls the electron orbit
topology through magnetic mirroring, curvature, and grad-$B$ drifts.
The magnetic part of the Lorentz force performs no work. The associated
energy conversion therefore represents electric-field work on the
drift currents, rather than direct work by the magnetic field. For the
grad-$B$ and curvature-drift channels, we distinguish the signed net
power from the positive, or gross-heating, contribution:
\begin{equation}
W_{\alpha}^{\rm net}(t)
=
\int_V
\mathbf{J}_{e,\alpha}\cdot\mathbf{E}\,
{\rm d}V ,
\qquad
W_{\alpha}^{(+)}(t)
=
\int_V
\left[
\mathbf{J}_{e,\alpha}\cdot\mathbf{E}
\right]_{+}
{\rm d}V ,
\end{equation}
where $\alpha\in\{\nabla B,\mathrm{curv}\}$ and
$[x]_{+}=\max(x,0)$. The difference between
$W_{\alpha}^{(+)}$ and $W_{\alpha}^{\rm net}$ measures the magnitude of
the spatially integrated cooling contribution. Thus, a large positive
value of $W_{\alpha}^{(+)}$ does not necessarily imply a comparably
large net heating rate.

The profiles and time traces show that both magnetic-drift channels
contain substantial regions of positive and negative
$\mathbf{J}\cdot\mathbf{E}$. These contributions partially cancel when
integrated over the discharge. The cancellation is especially strong
for the curvature-drift term: despite its appreciable positive
contribution, its net contribution to the time-averaged electron power
is only
\begin{equation}
\frac{
\left\langle W_{\mathrm{curv}}^{\rm net}\right\rangle_t
}{
\left\langle W_e\right\rangle_t
}
=
2.6\%.
\end{equation}
The grad-$B$ term also exhibits significant cancellation, but retains a
substantial positive net contribution,
\begin{equation}
\frac{
\left\langle W_{\nabla B}^{\rm net}\right\rangle_t
}{
\left\langle W_e\right\rangle_t
}
=
30.4\%.
\end{equation}
By comparison, the collisional heating varies much more weakly in time
and contributes
\begin{equation}
\frac{
\left\langle W_{\mathrm{coll}}\right\rangle_t
}{
\left\langle W_e\right\rangle_t
}
=
46.4\%.
\end{equation}
Consequently, the collisional term provides the largest relatively
steady contribution among these three channels, whereas the magnetic
drifts are characterized by a much stronger local exchange between
heating and cooling.

For the electrons, the field-based $\int_V\mathbf{J}_e\cdot\mathbf{E}\,{\rm d}V$
and the particle kinetic-energy change of the implicit mover agree at
the ${\approx}20\%$ level; the channel fractions above are quoted
relative to the field-based value and carry this uncertainty. The gross
powers and the spatial structure of the channels are insensitive to
this difference.

A plausible orbit-level interpretation of this signed energy exchange
is provided by the coupling between the electron bounce motion and the
small-scale instability. The instability generates a rapidly varying
axial electric field
$E_z(z,\theta,t)$ and thereby opens an axial channel for energy
exchange between the field and the electrons. During a bounce orbit,
the electron guiding center is displaced azimuthally by the
$\mathbf{E}\times\mathbf{B}$ drift and by the magnetic drifts. The
outgoing and returning portions of the orbit therefore pass through
the same axial position at different azimuthal locations and generally
at different phases of the fluctuating field. In particular,
\begin{equation}
E_z
\left(
z,\theta_{\rm out},t_{\rm out}
\right)
\neq
E_z
\left(
z,\theta_{\rm ret},t_{\rm ret}
\right) .
\end{equation}
The axial electric force acting on the electron during the outward
trajectory is therefore not exactly reversed during the return
trajectory. As a result, the contributions
$q_e\int E_z\,{\rm d}z$ from the two legs need not cancel. Depending on
the phase of the instability, an electron may either gain energy from
or return energy to the electric field. When summed over the electron
population, this orbit-to-orbit asymmetry produces alternating positive
and negative contributions to $J_{ez}E_z$ and, consequently, strong
fluctuations of the total electron power. It also provides a physical
connection between the small-scale instability and the negative
electron power observed near the cathode. This interpretation is
consistent with the present diagnostics, although it should be
regarded as a working physical picture rather than a unique
identification of the underlying instability.

The global power balance is examined in
Fig.~\ref{fig:power-balance}(b). The electrical power delivered by the
external circuit is
\begin{equation}
P_{UI}(t)=U(t)I(t),
\end{equation}
whereas the total electrical power transferred to the charged
particles is
\begin{equation}
W_{\mathrm{tot}}(t)
=
W_e(t)+W_i(t)
=
\int_V
\left(
\mathbf{J}_e\cdot\mathbf{E}
+
\mathbf{J}_i\cdot\mathbf{E}
\right)
{\rm d}V .
\end{equation}
The agreement of the time-averaged quantities,
\begin{equation}
\left\langle P_{UI}\right\rangle_t
\simeq
\left\langle W_{\mathrm{tot}}\right\rangle_t ,
\end{equation}
demonstrates global power closure. The instantaneous
$W_{\mathrm{tot}}$ nevertheless undergoes pronounced excursions around
$P_{UI}$. Because the ion power $W_i$ remains nearly constant during
the sampled interval, these excursions originate predominantly from
the electron power $W_e$. Figure~\ref{fig:power-balance}(b) therefore
shows that the small-scale instability strongly modulates the
instantaneous net energy transferred to the electrons, while having
only a weak effect on the ion heating.

Selected directional contributions to the electron power are also
shown:
\begin{equation}
W_{e\theta}(t)
=
\int_V
J_{e\theta}E_\theta\,
{\rm d}V,
\qquad
W_{e\parallel}(t)
=
\int_V
J_{e\parallel}E_\parallel\,
{\rm d}V ,
\end{equation}
with
\begin{equation}
J_{e\parallel}
=
\mathbf{J}_e\cdot\mathbf{b},
\qquad
E_\parallel
=
\mathbf{E}\cdot\mathbf{b},
\qquad
\mathbf{b}
=
\frac{\mathbf{B}}{|\mathbf{B}|}.
\end{equation}
The positive azimuthal work $W_{e\theta}$ follows the temporal
variation of $W_e$, indicating that the non-axisymmetric electric
fields associated with the rotating spoke and the small-scale
fluctuations constitute an important electron-energization channel.
In contrast, $W_{e\parallel}$ remains negative on average, showing that
the field-aligned electron population returns energy to the electric
field. However, its positive and negative components can both be substantial, as seen in Figure~\ref{fig:power-balance} (a), and the result also exhibits a fluctuating/oscillatory behavior. In regions where the magnetic field has a substantial axial
projection, this negative and oscillatory field-aligned work also
reflects the instability-driven axial energy exchange described above.

The directional quantities $W_{e\theta}$ and $W_{e\parallel}$ and the
mechanism-based quantities
$W_{\nabla B}$, $W_{\mathrm{curv}}$, and $W_{\mathrm{coll}}$ are
complementary diagnostics rather than mutually exclusive terms.
They should therefore not be added together to form a power-closure
relation. The global validation is instead provided by the comparison
between $P_{UI}$ and $W_{\mathrm{tot}}$. The horizontal coordinate in
Fig.~\ref{fig:power-balance}(b) is the simulation-step index $n$,
corresponding to the physical time $t=n\Delta t$, with
$\Delta t=5\times10^{-11}\ {\rm s}$.

Overall, the results reveal a clear distinction between gross and net
electron energization. The small-scale instability drives intense,
rapidly alternating electron heating and cooling, producing the large
fluctuations of $W_e$ in Fig.~\ref{fig:power-balance}(b). Much of this
bidirectional energy exchange cancels after spatial and temporal
integration, particularly for the curvature-drift channel, whereas
the grad-$B$ drift retains a substantial positive net contribution.

\subsection{Azimuthal Mode Spectrum of Electron Energy Deposition and Spoke-Ionization Coupling}

\begin{figure}[htbp]
  \centering
  \includegraphics[width=0.92\textwidth]{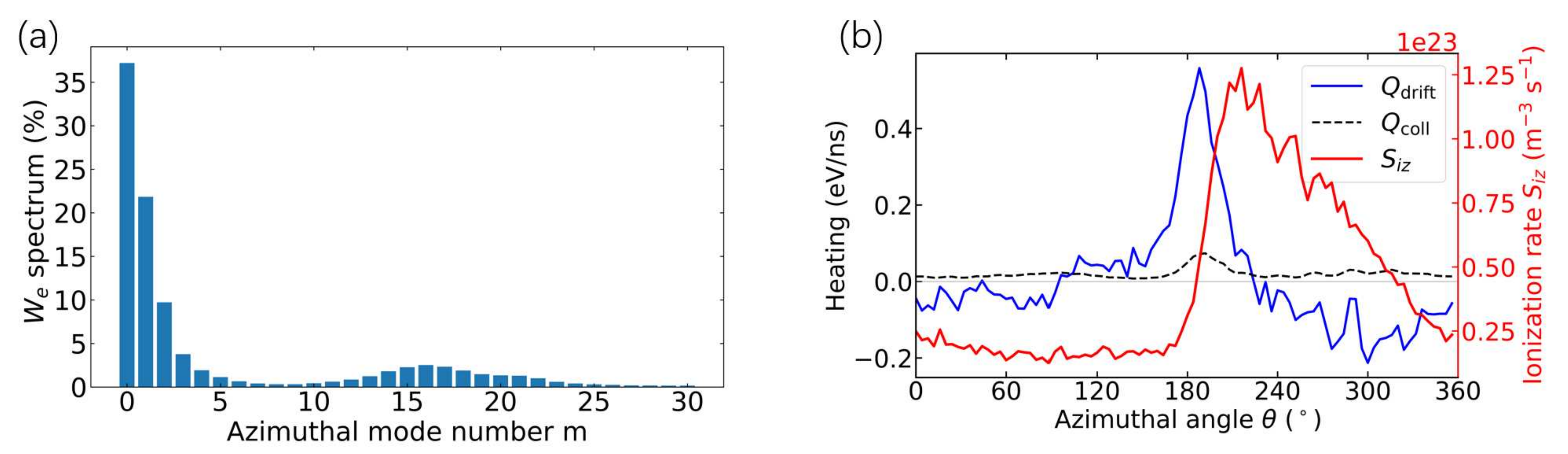}
  \caption{
    Azimuthal spatial spectrum of electron energy deposition and the azimuthal correlation between electron heating mechanisms and ionization rate.
    (a) Fourier-decomposed azimuthal mode spectrum of the electron energy deposition \( W_e \);
    (b) Azimuthal profiles of drift heating \( Q_\mathrm{drift} \), collisional heating \( Q_\mathrm{coll} \), and local ionization rate \( S_{iz} \) within a torus of major radius 1.2~cm and minor
radius 1~mm, 3~mm above the cathode. 
  }
  \label{fig:mode_spectrum_heating}
\end{figure}

Figure \ref{fig:mode_spectrum_heating} characterizes the azimuthal structure of electron energy deposition and its causal linkage to ionization production in the rotating spoke regime.

Panel (a) presents the azimuthal spatial spectrum of the numerically resolved electron energy deposition \( W_e \). Each azimuthal harmonic component is obtained via Fourier decomposition of the azimuthal profile \( W_e(r,z,\theta) \) at a given radial–axial location, defined as:
\begin{equation}
W_{e,m}(r,z)=\frac{1}{2\pi}\int_{0}^{2\pi}
W_e(r,z,\theta)e^{-im\theta}\,d\theta .
\end{equation}
The spectral share of each mode is quantified by its fractional contribution to the total azimuthal spatial modulation. As shown in the figure, the spectrum features a dominant low-order mode peak and a secondary broad peak at intermediate mode numbers. Collectively, non-axisymmetric components spanning \( m=1 \) to \( m=22 \) account for approximately 59.0\% of the total spatial modulation. Specifically, the low-mode band \( m=1\mbox{--}3 \) and the intermediate-mode band \( m=12\mbox{--}22 \) contribute 35.3\% and 18.0\% of the modulation, respectively. We emphasize that these fractions describe the strength of azimuthal periodic modulation and the associated alternating heating/cooling pattern, rather than the net power fraction in the global discharge energy balance.

Panel (b) illustrates the azimuthal evolution of two distinct electron heating channels alongside the local ionization rate. The blue solid line represents drift heating \( Q_\mathrm{drift} \) originating from collisionless field–particle interactions in the spoke electric field, the black dashed line represents collisional heating \( Q_\mathrm{coll} \), and the red solid line shows the ionization rate \( S_{iz} \) on the right-hand axis. All three quantities exhibit clear periodic modulation tied to the rotating spoke structure. A distinct phase shift is observed between the heating and ionization profiles: the peak of \( Q_\mathrm{drift} \) precedes the peak of \( S_{iz} \) in the azimuthal direction, establishing a causal sequence in which electrons are first energized by the spoke electric field, and the elevated electron temperature subsequently drives enhanced ionization. In contrast, \( Q_\mathrm{coll} \) remains nearly constant across the full azimuthal range with negligible modulation. This comparison demonstrates that the azimuthal modulation of ionization is sustained primarily by collisionless drift heating from the spoke, rather than by collisional energy transfer.

\subsection{Field-Line Resolved Cross-Field Transport and Non-Isothermal Electrons}

\begin{figure}[htbp]
  \centering
  \includegraphics[width=0.92\textwidth]{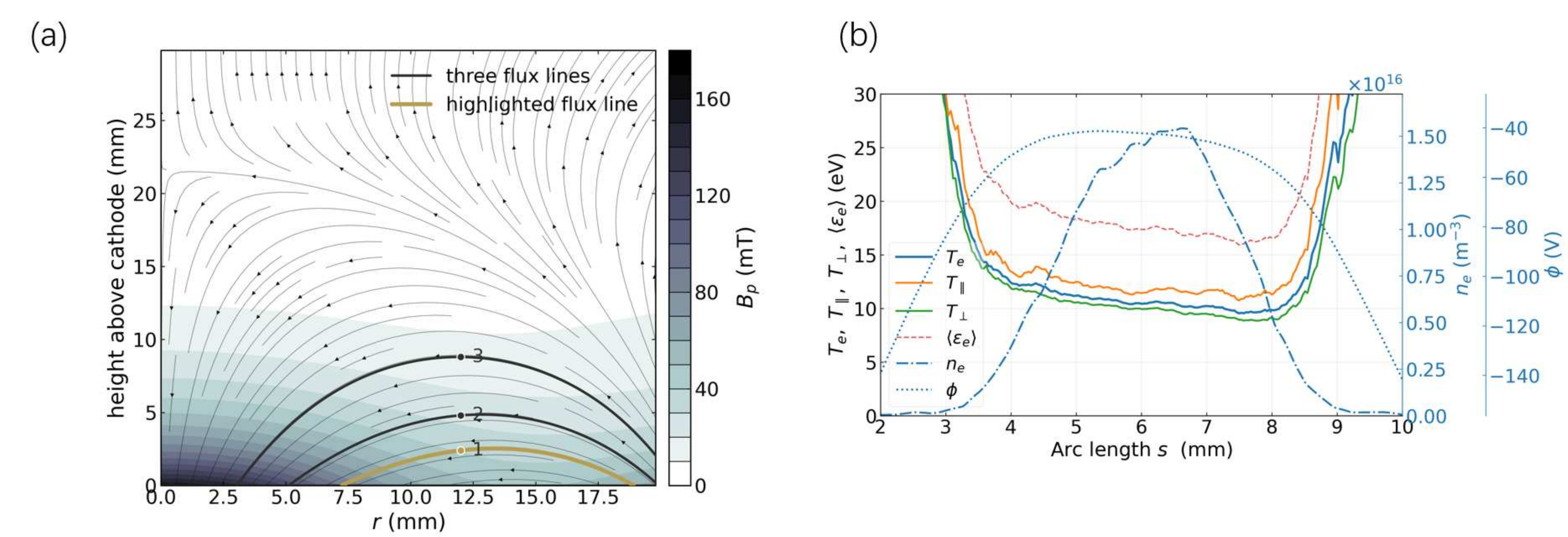}
  \caption{
    Selection of magnetic flux lines and electron property distributions along a highlighted field line.
    (a) Poloidal magnetic field \( B_p \) contour with three representative flux lines selected at different axial heights: line 1 (\( r=12\ \mathrm{mm}, z=2.6\ \mathrm{mm} \), highlighted), line 2 (\( r=12\ \mathrm{mm}, z=5\ \mathrm{mm} \), the same as the paper case), and line 3 (\( r=12\ \mathrm{mm}, z=9\ \mathrm{mm} \)). The fractional fluxes passing through the \( m=1\text{--}3 \), \( m=4\text{--}11 \), \( m=12\text{--}22 \), and \( m>22 \) energy bins, together with the total flux, are listed in the table below.  
(b) Profiles of electron temperature components, average electron energy, electron density, and electric potential along the highlighted flux line in (a) as functions of arc length \( s \).
  }
  \label{fig:field_line_profiles}
\end{figure}

To investigate the radial dependence of azimuthal perturbations and their contribution to cross-field electron transport, we select three representative magnetic flux lines at different axial heights above the cathode, as displayed in Fig. \ref{fig:field_line_profiles}(a). The background contour shows the poloidal magnetic field \( B_p \) distribution; black lines mark the three selected flux lines, and the yellow line highlights the lowest flux line near the cathode sheath. The three flux lines are numbered 1, 2, and 3 from the cathode upward, corresponding to the near-sheath region, the intermediate transition region, and the bulk plasma region, respectively.

We decompose the cross-field electron flux on each flux line into contributions from different azimuthal mode bands; the results are summarized in Table \ref{tab:cross_field_mode}.

\begin{table}[htbp]
  \centering
  \caption{Azimuthal mode decomposition of cross-field electron flux for the three selected flux lines. The last column lists the magnitude of the total cross-field electron flux.}
  \label{tab:cross_field_mode}
  \renewcommand{\arraystretch}{1.2}
  \begin{tabular}{lccccc}
    \hline
    Field line (coordinates) & \( m=1\mbox{--}3 \) & \( m=4\mbox{--}11 \) & \( m=12\mbox{--}22 \) & \( m>22 \) & Total flux (\(\mathrm{m^{-2}s^{-1}}\)) \\
    \hline
    1 (\( r{=}12\ \mathrm{mm}, z{=}2.6\ \mathrm{mm} \)) & 33.28\% & 16.01\% & 50.00\% & 0.71\% & \( 7.84\times10^{20} \) \\
    2 (\( r{=}12\ \mathrm{mm}, z{=}5\ \mathrm{mm} \), paper case) & 63.73\% & 11.97\% & 23.15\% & 1.14\% & \( 2.32\times10^{21} \) \\
    3 (\( r{=}12\ \mathrm{mm}, z{=}9\ \mathrm{mm} \)) & 96.68\% & $-0.69$\% & 3.72\% & 0.30\% & \( 4.07\times10^{21} \) \\
    \hline
  \end{tabular}
\end{table}

A clear crossover of the dominant mode is observed as one moves from the cathode sheath into the bulk plasma. Near the cathode sheath (field line 1), high-\( m \) perturbations in the range \( m=12\mbox{--}22 \) dominate the cross-field flux, contributing 50.0\% of the total. At the intermediate position (field line 2), low-\( m \) modes (\( m=1\mbox{--}3 \)) become the largest contributor at 63.7\%, while the high-\( m \) share drops to 23.2\%. Deep in the bulk plasma (field line 3), cross-field transport is overwhelmingly dominated by the large-scale spoke mode (\( m=1\mbox{--}3 \)), which accounts for 96.7\% of the total flux. This evolution indicates that small-scale, high-frequency azimuthal turbulence governs cross-field transport in the near-cathode sheath region, whereas the large-scale \( m\sim1 \) spoke structure dominates transport in the bulk plasma, with the transition of dominance occurring at the sheath–bulk interface.
We further validate the mode decomposition by comparing the total cross-field electron flux directly sampled from the PIC simulation (\( \Gamma_\mathrm{PIC} \)) with the flux reconstructed from all non-axisymmetric Fourier components (\( \Gamma_{m\geq1} \)). Because the flux surface is represented on a finite grid and is not exactly field-aligned at the discrete level, some parallel flux contribution inevitably leaks into the nominal cross-field count. This numerical error primarily arises in regions where the magnetic field lines are strongly tilted. Nevertheless, the overall discrepancy remains modest, corresponding to an 18.1\% relative difference. This level of consistency further corroborates that cross-field electron transport is predominantly driven by the large-scale spoke and small-scale azimuthal instabilities.

We further analyze the electron temperature and energy distribution along the highlighted flux line (line 1), with results plotted in Fig. \ref{fig:field_line_profiles}(b). The profiles show parallel electron temperature \( T_\parallel \), perpendicular electron temperature \( T_\perp \), average electron energy \( \langle \varepsilon_e \rangle \), electron density \( n_e \), and electric potential \( \phi \) as functions of arc length \( s \) along the flux line.

Recently, Suazo Betancourt \textit{et al.}~\cite{suazo2024} experimentally demonstrated via laser Thomson scattering that magnetic field lines in Hall effect thrusters are not isothermal, contradicting the long-standing assumption of isothermal field lines adopted in many fluid models. Our simulation results are quantitatively consistent with this experimental finding. Even along the central segment of the flux line, far from the end boundaries, the average electron energy varies between 16 eV and 20 eV, exhibiting clear non-isothermal behavior. Additionally, the average electron energy rises sharply at both ends of the flux line, corresponding to the near-cathode and near-anode boundary regions. This observed non-isothermal nature of magnetic field lines provides direct support for our analysis of radial current \( j_r \): parallel temperature gradients along field lines can drive substantial field-aligned currents, which couple to the radial current component and play a critical role in cross-field transport and spoke formation.

\section{Supplemental movie}

Supplemental Movie~1 shows the simultaneous evolution of the three-dimensional
electron density and the field-aligned electron-current density.  In the
saturated state, the broad $m=1$ density spoke propagates in the
$-\bm{E}\times\bm{B}$ direction, whereas the fine-scale current-filament
pattern propagates in the $+\bm{E}\times\bm{B}$ direction.  

Supplemental Movie~2 displays the dynamic variation of the current and the streamlines over time.
\FloatBarrier